\documentclass{article}
\usepackage{comment}
\usepackage[utf8]{inputenc}
\usepackage{amsthm}
\usepackage{graphicx}
\usepackage{subcaption}
\usepackage[T1]{fontenc}
\usepackage[utf8]{inputenc}
\usepackage{authblk}
\usepackage{tikz}
\usepackage{mathtools}
\usepackage[a4paper, total={6in, 10in}]{geometry}
\usepackage{nameref}
\usepackage{varioref}
\usepackage{hyperref}
\usepackage{cleveref}
\usepackage{soul}

\usepackage[edges]{forest}
\usepackage{adjustbox}
\usepackage[utf8]{inputenc} 
\usepackage[T1]{fontenc} 
\usepackage{multimedia}

\providecommand{\keywords}[1]{\textbf{Keywords -} #1}

\title{Network Security Modeling using NetFlow Data: Detecting Botnet attacks in IP Traffic}
\author[1]{Ganesh Subramaniam}
\author[2]{Huan Chen}
\author[3]{Ravi Varadhan}
\author[1]{Robert Archibald}
\affil[1]{Data Science \& AI Research, AT\&T Chief Data Office}
\affil[2]{Department of Biostatistics, Johns Hopkins Bloomberg School of Public Health}
\affil[3]{Division of Biostatistics and Bioinformatics, Department of Oncology, Johns Hopkins University}

\begin{document}


\title{Network Security Modeling using NetFlow Data: Detecting Botnet attacks in IP Traffic}


\maketitle

\begin{abstract}
Cybersecurity, security monitoring of malicious events in IP traffic, is an important field largely unexplored by statisticians. Computer scientists have made significant  contributions in this area using statistical anomaly detection and other supervised learning methods to detect specific malicious events. In this research, we investigate the detection of botnet command and control (C\&C) hosts in massive IP traffic. We use the NetFlow data, the industry standard for monitoring of IP traffic for exploratory analysis and extracting new features. Using statistical as well as deep learning models, we develop a statistical intrusion detection system (SIDS) to predict traffic traces identified with malicious attacks. Employing interpretative machine learning techniques, botnet traffic signatures are derived. These models successfully detected botnet C\&C hosts and compromised devices. The results were validated by matching predictions to existing blacklists of published malicious IP addresses.
\end{abstract}



\section{Introduction}

Security monitoring of Internet Protocol (IP) traﬃc is an important problem and growing in prominence. This is a result of both growing internet traﬃc and a wide variety of devices connecting to the internet. Along with this growth is the increase in malicious activity that can harm both individual devices as well as carrier networks. Therefore, it is important to monitor this IP traﬃc for malicious activity and ﬂag anomalous external IP addresses that may be causing or directing this activity through communications with internal devices on a real time basis. 

In network security, there are a large number of challenging statistical problems. For example, there is ongoing research associated with identification of various malicious events like scanning, password guessing, DDoS attacks, looking for malware and different spams attacks. The focus of this paper is on the detection of botnet attacks, specifically identifying host IP addresses (also known as C2 or "Command and Control") that send instructions to the infected bots (infected devices) on the nature of the attack to be perpetrated.

Reviewing the literature in the network security area, we observed that the current trend is device-centric, i.e.,\ analysis of the device’s traffic to determine whether it contains malicious activity. Evangelou and Adams, \cite{Evangelou2016PredictabilityON} construct a predictive model based on Regression Trees that models individual device behaviour that depends on features constructed from observed historic NetFlow data. By contrast, we are formulating a host-centric analysis where we are looking for external host IPs (that the devices are connecting to) that are possibly acting in a malicious way, particularly as the command and control server (C2) for a botnet. While a Bot device may have only a portion of its traffic be malicious among its benign traffic, the Host / C2 will have the majority of its traffic involved in the malicious activity and therefore have a stronger signature. Our analysis is looking for these signatures. Once the malicious Host is identified, the associated devices can be reviewed for infection.

Most of the current work in botnet detection has come from the computer science community. For example, Tegeler, et al, \cite{tegeler2012botfinder} used ﬂow-based methods to detect botnets. Choi et al, \cite{choi2009botgad} detected botnet traﬃc by capturing group activities in network traﬃc. Clustering is another approach taken by researchers to detect botnets using ﬂow based features. Karasaridis, et al, \cite{karasaridis2007wide} developed a K-means based method that employs scalable non-intrusive algorithms that analyze vast amounts of summary traﬃc data. Statisticians have a lot of potential to oﬀer new advanced analytic frameworks and techniques for botnet detection in network security related problems.

We present a statistical pipeline to model the IP network traﬃc for a given day using the NetFlow data and to detect botnet attacks without deep packet inspection. We develop a statistical intrusion detection system (SIDS) to detect malicious traﬃc particularly related to botnet attacks. Once a malicious host IP address is identiﬁed there are several actions that can be taken. Investigations can be conducted to understand the nature of the activity then mitigate or block the attack. Impacted devices, i.e., the devices being infected, attacked or abused by the external host, can be identiﬁed and cleaned. The security data oﬀers a lot of scope for applying classical statistical data exploratory techniques, machine learning models and the new deep learning methods to predict whether an IP is viewed as malicious.

\subsection{\textsf{NetFlow Data}}

This work uses the NetFlow data from the carrier’s IP traffic for all the modeling and analysis. NetFlow is a network protocol developed by Cisco for collecting, analyzing and monitoring of the packet capture data. It is the fundamental data for characterizing IP traﬃc, comprising source and destination IP addresses, packets and bytes transferred, duration and IP protocol number used. But there are other data components as well, e.g., data from HTTP log ﬁles and DNS requests. By limiting ourselves to the NetFlow data, we can use this relatively available source to create a funnel of highly probable IPs for further, more intensive investigation.

There are a number of data challenges that increase the complexity associated with the analysis and modeling of the security data. 

\begin{itemize}
  \item The ﬁrst issue is the sheer volume of the data. Netflow data for a single day across all classes of IP traﬃc could run into several terabytes of data. It calls for eﬃcient storage, processing and computing. Due to the prohibitive size, even with the use of big data platforms there is usually a short history available. 
  \item Establishing ground truth is hard. Training data with both malicious and benign labels is difficult to develop. The process of conﬁrming an IP address as a bad actor requires processes that are expensive in terms of time and eﬀort. It may require further investigations like deep packet inspection (DPI), processing that inspects, in detail, the content of the IP packets looking for malicious code.  There have been some attempts at developing samples to work with such as the CTU project \cite{ctu13} that developed samples around a number of known malwares.
  \item The statistics problem, from a modeling standpoint, is the imbalance of the classes and an Unknown class, i.e. a few known Bad vs many Unknown. The imbalance is due to the limited availability of labeled IP addresses associated with botnet traﬃc, i.e., known “bad actors”. The second class is not Good, i.e., known not Bad, but rather the remaining traﬃc is largely unknown, i.e. contaminated with both good and bad.
\end{itemize}
 
\begin{table}[!htbp]
\begin{center}
\begin{tabular}{ |l|l| }
  \hline
  \multicolumn{2}{|c|}{NetFlow Data Fields} \\
  \hline
  1. & Source IP address \\
  2. & Destination IP address\\
  3. & Source port \\
  4. & Destination port\\
  5. & Bytes transferred    \\
  6. & Packets transferred    \\
  7. & Start Time \\
  8. & End Time \\
  9. & IP Protocol number\\
  10. & Flag \\
   \hline
\end{tabular}
  \caption{NetFlow Data}
  \label{tab:netflowtab}
  \end{center}
\end{table}
The NetFlow data has a limited number of attributes, shown in Table~\ref{tab:netflowtab}. We need to extract additional relevant statistical features that can be used in a predictive model to classify bot traﬃc vs normal traﬃc  (see section 1.4)
 
\subsection{\textsf{What is a Botnet?}}
 
In recent years botnets have emerged as one of the biggest threats to network security among all types of malware families as they have the ability to constantly change their attack mechanism in scale and complexity \cite{silva2013botnets}. A botnet is a network of compromised devices called bots and one or more Command \& Control (C\&C or C2) servers. Generally speaking, the bots could be a PC, a server, an Internet of Things (IoT) device or any machine with access to the Internet. In this type of threat, the botmaster, which is the orchestrator, authors a malware that operates on each bot. Devices are infected with the malware in several ways such as: "drive by downloads" which refers to the (unintentional) download of malware as a result of just visiting a site or from infected emails. The botnet control system, or C2 for Command \& Control (C\&C) server, is the mechanism used by the botmaster to send commands and code updates to bots which then conduct the attacks. Due to the prevalence of firewalls, the botmaster cannot contact devices directly. Typically, the bot malware has instructions to contact the C2 to establish the communications and to receive instructions on any attacks to be perpetrated. The nature of the attacks vary in scale and sophistication. Examples of attacks by botnets include transmitting malware, using the bots to perform diﬀerent illegal activities, e.g.,\ spamming, phishing, or stealing conﬁdential information, and orchestrating various network attacks (e.g.,\ DDoS – Distributed Denial of Service). 

\subsection{\textsf{Approach}}

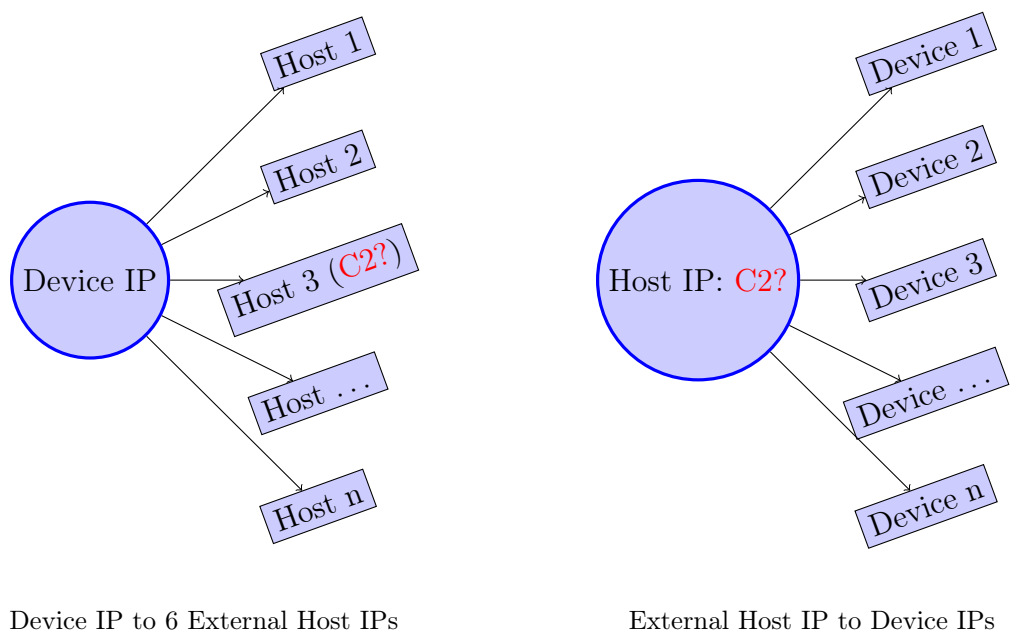
\begin{figure}[!ht]
\begin{center}
\usetikzlibrary{shapes.geometric}
\begin{tikzpicture}[fill=blue!20]
\path (-5,0) node(a) [circle,draw=blue,very thick,fill] { \large Device IP }
      (-2,3) node(b) [rectangle,rotate=20,draw,fill] {\large Host 1}
      (-2,1.5) node(c) [rectangle,rotate=20,draw,fill] {\large Host 2}
      (-2,0) node(d) [rectangle,rotate=20,draw,fill] {\large Host 3 (\textcolor{red}{C2?})}
      (-2,-1.5) node(e) [rectangle,rotate=20,draw,fill] {\large Host $\dots$}
      (-2,-3) node(f) [rectangle,rotate=20,draw,fill] {\large Host n}
      (3,0) node(g) [circle,draw=blue,very thick,fill] {\large Host IP: \textcolor{red}{C2?}}
      (6,3) node(h) [rectangle,rotate=20,draw,fill] {\large Device 1}
      (6,1.5) node(i) [rectangle,rotate=20,draw,fill] {\large Device 2}
      (6,0) node(j) [rectangle,rotate=20,draw,fill] {\large Device 3}
      (6,-1.5) node(k) [rectangle,rotate=20,draw,fill] {\large Device $\dots$}
      (6,-3) node(l) [rectangle,rotate=20,draw,fill] {\large Device n}
      (-3.5,-4.5) node(m) [rectangle] { Device IP to 
6
 External Host IPs}
      (4.5,-4.5) node(n) [rectangle] { External Host IP to Device IPs};

\draw[->] (a) -- (b);   
\draw[->] (a) -- (c);
\draw[->] (a) -- (d);
\draw[->] (a) -- (e);
\draw[->] (a) -- (f);
\draw[->] (g) -- (h);   
\draw[->] (g) -- (i);
\draw[->] (g) -- (j);
\draw[->] (g) -- (k);
\draw[->] (g) -- (l);
\end{tikzpicture}
\end{center}
\caption{Device Centric vs. C2-Centric Traffic Flow}
\label{fig:DevVsHostFlow}
\end{figure}

We took a C2-centric view of the data. As noted above, the most common approach to identifying botnets is to look at individual devices and analyze their traffic with various hosts. This means analyzing each
of the device’s connections, as shown in the left panel of Figure~\ref{fig:DevVsHostFlow}, for possible malicious traffic. However, the connection with the C2 may not look significantly different than the other benign traffic for that device or be a small portion of its traffic. Therefore, each connection must be analyzed individually. 

In our approach, we analyze the external host for C2 behavior. The right panel of Figure~\ref{fig:DevVsHostFlow} shows the traffic between one external host IP address and several devices that are internal to a carrier network, each having a distinct IP address. Thus, for each host IP address (an external host), we aggregate the device traffic. The question is which host IP address has traffic that looks like a \textit{botnet command and control (C2)} pattern? Most of the C2’s traffic should be botnet related and it should be doing this with a large number of bot devices. Therefore, we can look for the C2 signature as the predominant traffic pattern over all its paired devices. We call this the C2-centric approach . It allows for more aggregation and fewer cases that have to be analyzed, thereby improving its scalability.


We construct features for each host IP from the NetFow data traffic between the host and all of its associated device IPs (see right panel in Figure~\ref{fig:DevVsHostFlow}). Features include the number of flows, the number of unique devices, average number of packets, average duration, etc. (see section 1.4). We develop signatures, based on these features, for host IPs in known botnet families. We then model the constructed data as one observation per Host IP using the signature features.

\subsection{\textsf{Statistical Feature Analysis} }

The first important step is the exploration of the feature space such that they sufficiently describe the NetFlow traffic as features are really the lens through which the machine learning model views the data.  The ability of the feature space to provide pertinent information is critical to the machine learning step, as the underlying assumption of these classification models is that feature characterization of the malicious botnet and benign NetFlow traffic have different distributions. For our exploratory analysis, we subset traffic associated with selected IP addresses that are from a known botnet family, in other words "live" botnet traffic (i.e.,\ C2 IP addresses). Using the flow data for these IP addresses that are associated with a known botnet family, we  attempt to uncover some of the main characteristics or signatures that differentiate normal traffic from botnet traffic. Most botnet families and other families (non-botnet) share some of the features or signatures developed in this research. We employ the C2-centric analysis to hand craft the features for the ML models. Instead of analyzing the traffic between every pair of nodes (IP addresses), we analyze the traffic between every individual external host IP address and the group of device IP addresses it contacts within the carrier network.  For a given external host IP, we can compute several flow based features associated with botnet traffic. Features are aggregated for traffic over a 24 hour window for a given day. We will show examples of some of these features. 

In our C2 Centric approach, most of our features are based on flows from the potential C2 IP to the device. Therefore, in the following discussion, we reference the potential C2/normal server as the SIP and the device/potential Bot as the DIP. Our objective is to classify the SIP as either a C2 or a “normal” server.

\begin{figure}[!ht]
    \centering
    \includegraphics[width=\linewidth]{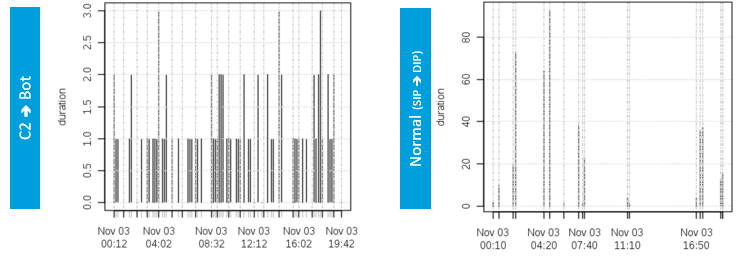}
    \caption{Flow Counts: C2 to Bot vs Normal Server to Device Traffic for the number of flows per SIP to all DIPs }
    \label{fig:flowCounts}
\end{figure}

The source IP flow count or the number of flows associated with each SIP can be an important indicator for  bot to C2 communication. Normal HTTP traffic generates a large number of flows in a few seconds while the flow counts associated with C2 to bots traffic are small and spread out over larger time intervals. This is done intentionally to maintain a low profile. The plot in Figure \ref{fig:flowCounts} demonstrates the flow count feature for normal and bot traffic.

\begin{figure}[!ht]
    \centering
    \includegraphics[width=\linewidth]{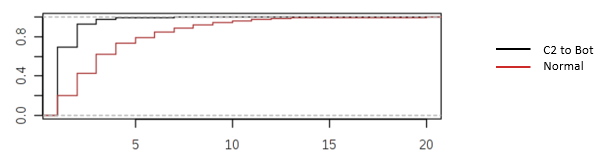}
    \caption{CDF for flow count: C2 to Bot vs Normal Server to Device Traffic for the number of flows per SIP to all DIPs}
    \label{fig:CDFflowCount}
\end{figure}

The plot of the cumulative distribution function (CDF) shown in Figure \ref{fig:CDFflowCount} is demonstrates the same concept, namely, the CDF of Flow Count shows values are $\le 2$ for 90\% of bot traffic, versus $\le 9$ for normal (per 10 min interval)

\begin{figure}[!ht]
    \centering
    \includegraphics[width=\linewidth]{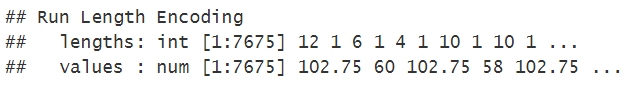}
    \caption{Example of Runlength.}
    \label{fig:runLength}
\end{figure}

Consider the all flows for a C2 to a Bot, in this case a C2 to several Bots. Let $P = \{p_1,p_2,p_3, ..., p_n \}$ where each $p_i$  represents packet size count in a single flow. Figure \ref{fig:runLength} shows an example of the first few run lengths. A run length is a streak of repeats of the same packet size (across all bots).

\begin{figure}[!ht]
    \centering
    \includegraphics[width=\linewidth]{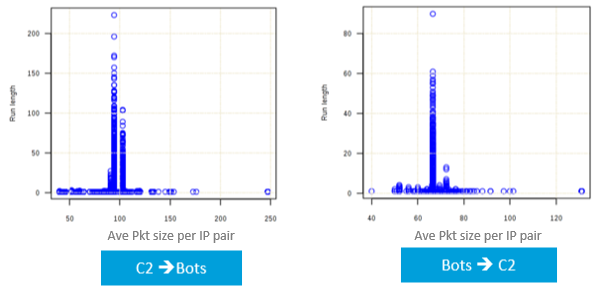}
    \caption{Example of dominant packet counts}
    \label{fig:packetCounts}
\end{figure}

For bot traffic, we observe similar packet counts, as shown in the plots in Figure \ref{fig:packetCounts}, where the majority of the large runs are for only one or two packet sizes. So these two packet sizes (102.75 and 98) dominate. The same is true in the reverse direction, namely, from bot to C2.

\begin{figure}[!ht]
    \centering
    \includegraphics[width=\linewidth]{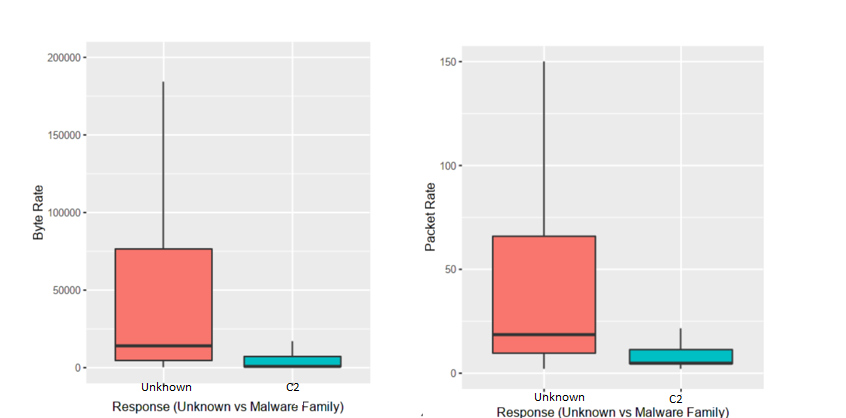}
    \caption{Byte per flow \& Packets per flow for Bot vs Normal traffic}
    \label{fig:bpr}
\end{figure}

So an important feature is to capture dominant packet sizes or counts. A similar phenomenon is observed for the bytes per packet ratio, where one or more ratios dominate. Another feature that stood out differentiating bot and normal traffic is the packets per flow and bytes per flow. These measures are computed by dividing the total number of packets or bytes for a SIP by the total number of flows observed for the SIP. The boxplots in Figure \ref{fig:bpr} show the differences in distribution for the known bot families and the unknown traffic. The median packets per flow is 5 vs 19 for the bot traffic vs normal traffic and the median bytes per flow is 1000 vs 14000.

The standard deviation of SIP packet count is another feature that is important in identifying bot traffic.  Typical IP traffic has variation in number of packet transferred, whereas for the bot traffic the standard deviation is quite low (typically less than 1). 

\begin{figure}[!ht]
    \centering
    \includegraphics[width=\linewidth]{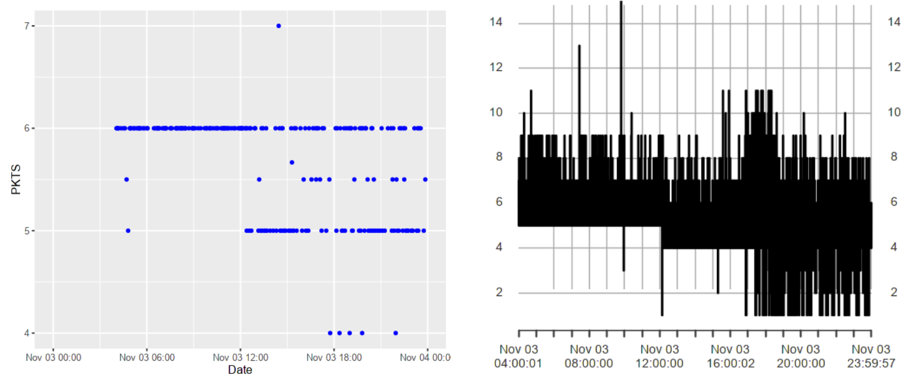}
    \caption{Examples of packet count time series}
    \label{fig:pcts}
\end{figure}

Also, if the average difference between start times of successive flows for a given SIP communicating with several devices is low that is another indicator for bot traffic. Over the course of a 24 hour window, if the source IP is sending packets to several devices on average, say at a rate equal to less than one second then that would be suspicious. The plot in Figure \ref{fig:pcts} shows the time series of packet count from the source IP to 612 devices. The mean time difference of successive start times for this time series can be an approximate measure of periodicity, which in this case is 0.68. Another feature of importance is who initiates the communication, the external IP or the internal device? Usually, the malware in the infected bot has the instructions to initiate contact with the C2. For this reason, if the internal device initiates the communication, then it is highly likely that the device is an infected bot and the external IP is a C2.  Further, if the source port (sport) used here is an HTTP port e.g.,\ 80 or 443, it is more likely to be a bot. Based on these kinds of exploratory analysis, two groups of features were identified, one group associated with the flow size and the other associated with beaconing activity. The flow size features are based on the assumption that traffic generated by bots are more uniform than traffic generated by normal users. For instance, if bots used fixed length commands these features will likely detect it. Below are some of the main flow size features that were considered for the model training:

\textbf{Flow size features:}

\begin{enumerate}
    \item Total number of bytes transferred: sum of bytes over all flows;
    \item Total number of packets transferred:  packet sum or sum of “number of packets” across all flows;
     \item Average bytes to packet ratio: This is the mean of the bytes to packet ratios over all flows also known as average length;
     \item No. of Dominant ratios: we compute the bytes per packet ratio for every flow. Then compute the unique number of ratios in 90\% of the flows ($90^{th}$ percentile). Normal traffic exhibits quite a bit of variation in the ratios computed. This feature is also computed for 65 and 75 percentiles of the flows;
     \item	Packets per flow: packets sum over the total flows, i.e.,\ ($\frac{\sum p_i }{\sum flows}$), provides average packets per flow;
    \item Bytes per flow: sum of bytes over the total flows, i.e.,\ ($\frac{\sum b_i }{\sum flows}$), provides average bytes per flow;
     \item	IQR Ratios: The inter-quartile distance is computed for the bytes per packet ratios for the sequence of flows;
    \item SD ratio: This is the standard deviation of the bytes to packet ratios;
    \item Dominant flow counts: We compute the number of flows in a 5 min interval over the 24-hour window. Then we compute the unique flow counts in 90\% of the time windows ($90^{th}$ percentile);
    \item Total duration: The duration is based on start time and end times for each flow, the aggregate duration for all flows for a given SIP is the total duration. This gives a measure of how long was the SIP communicating over the 24 hour window;
    \item DurMax/DurMed: Maximum duration and median duration over all flows;
    \item Sport/dport: Ports used for communication provides information on who initiated the contact, the host IP or the device;
    \item Flow frequency: Total number of flows. A few flows vs lots of flows either across the 24 hour interval or short bursts over some time periods;
    \item ctMax/ctMed: Maximum flow count and median flow count;
    \item Who initiated the connection? (Device or Host);
    \item Count of sport/dport: count for the most dominant ports;
    \item Unique no. of destination IPs (devices).
\end{enumerate}

\textbf{Beaconing features:} 

The malware downloaded by compromised internal devices or servers has a beaconing feature that involves the sending of short and routine communications to the C2 server, signaling that the internal infected computer is now available and listening for further instructions. We developed specific features to detect the presence of beaconing activity to confirm that the signaling is active. As an example, from the observed  sequence of source IP start times, the  \textit{inter-arrival times} are defined as the differences between start times of successive flows. If this is periodic, then beaconing signal is present, if this is random, then it’s not. The following features measure the beaconing signal effect:

\begin{enumerate}
   \item Periodicity for inter-arrival times -- start times of successive flows ($d_i = t_{i+1} - t_{i}, i=1,2,\dots$);
   \item	Time Gap: The average inter-arrival times difference between start times of successive flows ($\overline{d_i}_{i=1}^{n} = \sum \frac{d_i}{n}$);
    \item	SD number of packets: standard deviation for packet count, i.e.,\ number of packets by flow;
    \item Standard deviation of inter-arrival times (low values implies periodic).
\end{enumerate}
    
\section{Statistical modeling}

\subsection{Training ML Models for NetFlow Data}

Many researchers have applied machine learning (ML) techniques for botnet detection.  ML approaches use the IP traffic associated with a set of known IP addresses as training data and learn a
prediction function to classify an IP address as benign or malicious. The general principle is that the training data is labeled. The machine will “learn” from the labeled patterns to build the classifier and use it to predict class labels for new data. We considered several supervised learning algorithms including the random forest, gradient boosting, SVM, and LASSO for binary classification of the botnet traffic. Random Forests best handled the class imbalance as part of the model fitting process. Hence, the discussion of results will focus only on the random forest model. 

Random forest, introduced by Breiman \cite{breiman2001random}, has become one of the most popular out-of-the-box prediction tools for machine learning. This algorithm instead of using prediction from a single decision tree, grows several thousands of large trees, each built on a different bootstrap sample of the training data from which an aggregate prediction is calculated. The final prediction is a bagging or bootstrap aggregation where the trees are averaged. In order to classify new data, a majority vote is used, i.e.,\ the class predicted by the largest number of trees is chosen as the prediction. 

Borrowing notations from \cite{wager2016asymptotic} suppose that we have training examples $Z_i= (X_i,Y_i)$ for $i = 1,2 ..\dots n$, a test point $x$, for which prediction is sought, and a
regression tree predictor T which makes predictions $\hat{y} = T(x_i; Z_1,Z_2, ..\dots Z_n)$. We can then turn this tree $T$ into a random forest by averaging it
over $B$ random samples:

$$RF^{B}_{s}(x; Z_1,Z_2, ..\dots Z_n) = \frac{1}{B} \sum_{b=1}^{B}T(x; Z^{*}_{b_1},Z^{*}_{b_2}, ..\dots Z^{*}_{b_s}) \quad \mbox{for some} \quad s \le n,$$
where $\{Z^{*}_{b_1},Z^{*}_{b_2}, ..\dots Z^{*}_{b_s}\}$ 
form a uniformly drawn random subset of $\{Z_1,Z_2, ..\dots Z_n\}$. 
In the case of classification however, we take majority vote:
$$RF^{B}_{s}(x; Z_1,Z_2, ..\dots Z_n) = \mbox{majority vote}\{T(x))\}_{b=1}^{B}$$
The usual rule of thumb for the best sample size is a "bootstrap sample", a sample equal in size to the original data set, but selected with replacement, so some rows are not selected, and others are selected more than once.   

To ensure that the decision trees are not the same (or reduce correlation), random forest randomly selects a subset of the characteristics at each tree split. So instead of using all the variables, only a subset variables chosen at random will be used at every stage of splitting. This process decorrelates the resulting trees. The number of variables considered at every split of a node is a hyper-parameter that is estimated during the  training process. The reason that this algorithm works is that while individual trees based the bootstrap-resampled training data are considered weak learner, the ensemble process called bagging, which averages the independent trees to get the predictor, is considered a strong learner.

\subsection{Setting up training data for modeling}

The daily flow data is processed to extract the features discussed in the exploratory analysis. From the new processed data, the feature engineering process extracted close to 40 features for each source IP address. The source IP address could be associated with an external host IP or an internal device as the flow data captures conversations on both directions. The response column is assigned either “unknown” label or “malicious Family”. Labels were derived with the help of threat intelligence platform that maintains a list of confirmed IP addresses that belong to several malicious botnet families. We choose the active malware sample traffic traces that were observed in the network within a window of 30 days. These IP addresses that are related to the different botnet malware families are labeled “malicious family” label, and traffic associated with rest of the IP addresses are labeled “unknown”. We observed a significant imbalance in class labels as the list of IP addresses associated with the malicious families is very small compared to the entire traffic. This was one of the challenges, namely choosing the appropriate type of machine learning model that can accommodate this imbalance. To construct a training data set, the flow data was processed for one full month to create a training data set. About 1000s IP addresses from the "unknown" class were sampled for every day of the month. All traffic from IP addresses associated with "malicious" class traffic from the entire month were selected. This hand constructed training data had $\sim$ 17\%  malicious and 83\% unknown traffic.

\subsection{Unbalanced data}
Random forest by default uses several bootstrap samples from the original training data as a whole. However, if the training data has an unbalanced response, it presents a challenge for random forest classification. Here the samples are not uniformly distributed across the categories, but some of the categories have much larger or much smaller number of observations compared to the others. This leads to a bias in prediction towards more-common classes. The algorithm treats all classifications the same.  In this cases we are often most interested in correct classification of the rare class. In the NetFlow training data \ only $\sim$ 17\% of the traffic are associated with the botnet attack. To deal with this data imbalance, we use the \textit{balanced random forest} where a bootstrap sample is drawn from the minority class and then the same number of cases is drawn from the majority class.
This implementation is called \textit{down-sampling} since we down weight the sampling of the majority class (benign traffic) in the bootstrap sample. The training process used a "down-sampled" Random Forest where each tree is built from a bootstrap sample from the rare class ("malicious"), along with a sub-sample of the same size from the more prevalent class ("unknown").

\subsection{Results from the Random Forest Model}

The Random Forest model comes with a built-in validation process within the training process, eliminating the need for cross-validation or a separate test set to get an unbiased estimate of the test set error. At a tree level, as each tree is grown on its own bootstrap sample $Z^{*}_b$ it has its own test set called \textit{out-of-bag} (OOB) sample composed of observations that were not selected
into $Z^{*}_b$. From each bootstrap sample, the error rate for the observations left out of the bootstrap sample is monitored. This is called the \textit{out-of-bag} or OOB error rate. Using the OOB sample for each tree we can compute an unbiased estimate of our prediction error that is almost identical to the one obtained by $n$-fold cross-validation \cite{HastieTib2016} so this additional procedure is no longer necessary. 

\begin{figure}[!ht]
    \centering
    \includegraphics[width=\linewidth]{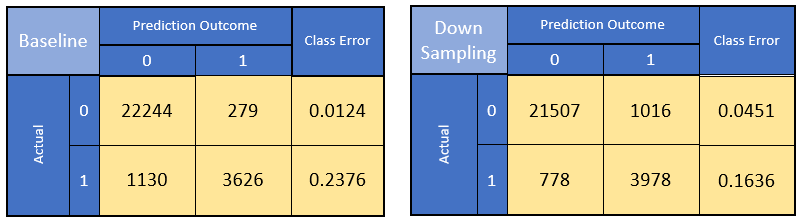}
    \caption{Confusion Matrix for model with down sampling vs baseline model ( without down sampling)}
    \label{fig:table_BRF}
\end{figure}

The modeling utilized 2500 trees. One of the key parameters in the Random Forest \textsf{R} package is the \textit{mtry} parameter that corresponds to the number of variables randomly sampled at each split. This was determined to be 10 based on cross-validation. The resulting OOB error is about 6.5\%. An average false negative rate of 16\% $(FN/(FN+TP))$ and false positive rate of 4.5\% $(FP/(FP+TN))$ were observed across the 2500 trees. We then validated the RF prediction model using 30 days of NetFlow traffic from a different month than the training data. The sample confusion matrix for the first day of the validation month is shown in Figure \ref{fig:confusionMatrix}. The plot in Figure \ref{fig:timeComp}, shows the daily accuracy, false positive rates and true positive rates for the entire validation month. The true positive rates fluctuates between 0.7 and 0.85 and seem to exhibit a downward trend. The range for accuracy and false positive rates are quite narrow. The down-sampling approach improved the prediction accuracy of the rare category by $\sim$ 30\% (Table in Figure \ref{fig:table_BRF}) with the added benefit of improved computation times for large data.

The IP addresses that are predicted to be malicious were further investigated for analysis. These are IPs with class label “unknown” that the model classified as potentially suspicious or malicious IPs. To evaluate the model-identified malicious IP addresses, we compare or match them with well-known blacklisted IP addresses available from several threat intelligence sharing platforms. The matched IP addresses from the daily lists provided additional threat information on these IPs, e.g., ISP that owns the IP address, the country of origin, types of attacks perpetrated etc. which gets published in a dashboard for security analysts.

\subsection{Interpretability of Machine Learning Algorithms}

Predictions based on complex Machine Learning models have been used in
several domains such as finance,  advertising, marketing, and medicine. Classical statistical learning algorithms like linear and logistic regression models are easy to interpret and there is a lot of history and practice in their application.  On the other hand, random forest, gradient boosting, deep learning and other learning algorithms have recently proven to be powerful classifiers, quite often surpassing the performance of the classical regression models.  However, a major challenge is that their prediction structure is not transparent. It is difficult for the domain experts to learn from these models. Researchers \cite{cShawiRadwa2019}, \cite{WangKaushal2020} 
often have to make the trade off between model interpretability and prediction performance. Therefore, interpretative machine learning has recently become an important area of research.  

Random Forest provides variable importance for features to help identify the strongest predictors, but it provides no insight into the functional relationship between the features and the model predictions. The first approach to assessing importance of predictors in a random forest was proposed by Breiman in his paper \cite{breiman2001random} and it is widely used even today. It adopts the idea that if a variable plays a role in predicting our response, then perturbing it on the out of bag (OOB)
sample should decrease prediction accuracy of the forest on this sample. Therefore, taking
each variable in turn, one can perturb its values and calculate the resulting average decrease of predictive accuracy of the trees -- such a measure is sometimes referred to as \textit{variable importance} (VIMP). While this is useful, it still does not provide adequate insight into the prediction structure of the algorithm and the inter-relationship between the predictors.

Ideally, what we would like to know is the following:
\begin{itemize}
    \item How do changes in a particular feature variable impact model's prediction?
    \item What is the relationship between the outcome and the individual features?
    \item Are these relationships approximately linear, monotonic or more complex?
    \item Are their strong interactions between predictors?
\end{itemize}

\begin{figure}%
    \centering
    \includegraphics[width=7cm]{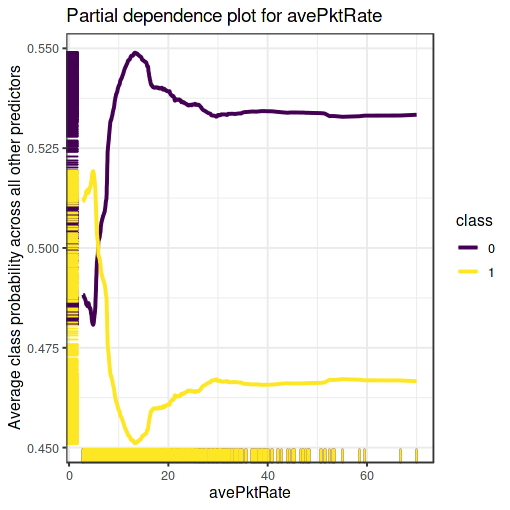} %
    \qquad
    \includegraphics[width=7cm]{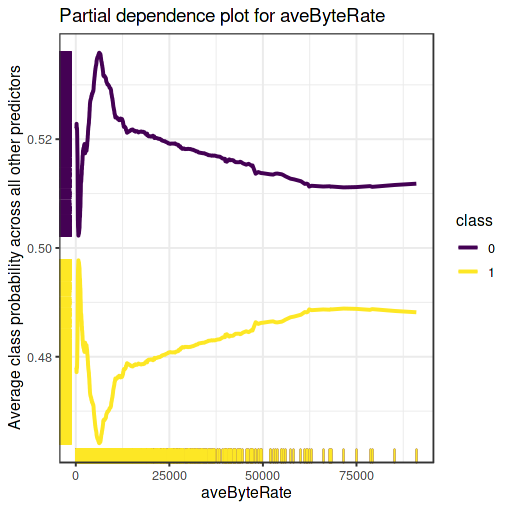} %
    \caption{Partial dependence plot for average packets per flow and bytes per flow.}%
    \label{fig:partialDependence}%
\end{figure}

Having this knowledge aids in better interpretation and creates more trust in the prediction algorithm since domain experts are unlikely to trust algorithms that are opaque. The partial dependence plot \cite{friedman2001greedy} or PDP is one of the many model agnostic techniques that helps achieves this goal. The partial dependence function for the model is defined as
$$\hat{f}_{xS}(x_S,x_C) = E_{xC}[\hat{f}(x_S,x_C)]= \int \hat{f}(x_S,x_C)dP(x_C)          $$
The $x_S$ are the features for which the partial dependence function should be plotted and $x_C$ are the other features used in the machine learning model $\hat{f}$. We produce the partial dependent plot by setting all but one feature to fixed values and compute average predictions by varying the remaining features throughout the complete range. This basically gives you the "average" trend of that variable integrating out all others in the model. The PDPs for the top two variables identified by the Random Forest model are shown in Figure ~\ref{fig:partialDependence}.

The x-axis represents the values of the feature variable and the y-axis is the average predicted effect. It shows the functional relationship between the prediction and the feature variable. For the "malicious" class, as average packets per flow increases, the probability of IP being classified as "malicious" decreases. Similarly, for the Unknown class, as average packets per flow increases, the probability of IP being classified as "unknown" increases. The value 16 for average packets per flow is the maximum separation. The important take away from this plot is that this offers a signature for individual features for identifying botnet traffic in NetFlow data (although being a single feature snapshot). 

\subsection{Importance of variables in forests}

As discussed in the last section, the ML models for features generally provide a relative importance, a value for each variable that indicates the strength of the relationship between each input and model's predictions. Ishwaran et al. \cite{Ishwaran} described a new paradigm for forest variable selection based on a tree-based concept termed \textit{minimal depth}. Paluszynska (2017) uses minimal depth to develop a multi-way importance measure for the Random Forest and is implemented in the \textsf{R} package \textit{randomForestExplainer}  \cite{paluszynska2017RFE}. Minimal depth is a measure of the distance of a variable relative to the root of the tree for directly assessing the predictiveness of a variable.

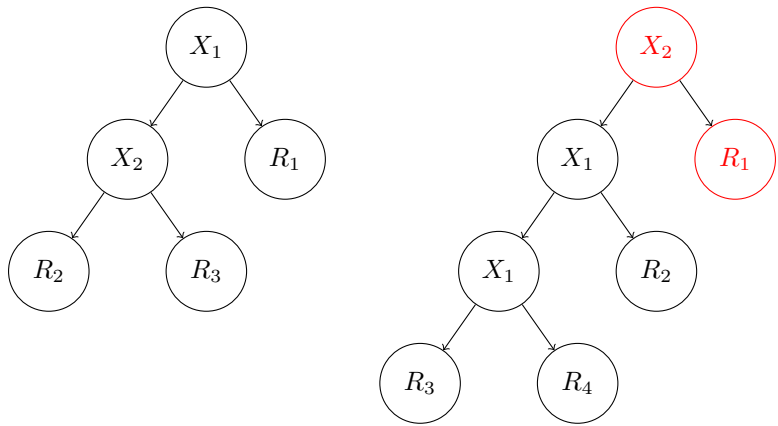
\begin{figure}
 \centering  
\begin{adjustbox}{valign=t}
\begin{forest}
for tree={
    grow=south,
    circle, draw, minimum size=7ex, color=black,inner sep=1pt,edge+=->,
    s sep=10mm
        }
[\textcolor{black}{$X_1$}
    [\textcolor{black}{$X_2$}
        [\textcolor{black}{$R_2$}]
        [\textcolor{black}{$R_3$}]
    ]
    [\textcolor{black}{$R_1$}]
    ]
\end{forest}
\end{adjustbox}\qquad
\begin{adjustbox}{valign=t}
\begin{forest}
for tree={
    grow=south,
    circle, draw, minimum size=7ex, color=black,inner sep=1pt,edge+=->,
    s sep=10mm
        }
[\textcolor{red}{$X_2$},color={red}
    [\textcolor{black}{$X_1$}
        [\textcolor{black}{$X_1$}
            [\textcolor{black}{$R_3$}]
            [\textcolor{black}{$R_4$}]
            ]
        [\textcolor{black}{$R_2$}]
    ]
    [\textcolor{red}{$R_1$},color={red}]
    ]
\end{forest}
\end{adjustbox}
\caption{Illustration of the concept of maximal subtrees. Maximal $X_1$-subtrees are highlighted in black. In the first tree, $X_1$ splits the root so the maximal $X_1$-subtree is the whole tree. In the second tree the maximal $X_1$-subtree contains an $X_1$-subtree that is not maximal. Source: Paluszynska (2017).   }
    \label{fig:maximalTree}
\end{figure}

This idea can be formulated precisely in terms of a maximal subtree (illustrated in Figure ~\ref{fig:maximalTree}). The maximal subtree for a variable v is the largest subtree whose root node is split using v (i.e., no parent node of the subtree is split using v). The shortest distance from the root of the tree to the root of the closest maximal subtree of v is the minimal depth of v.
A smaller value corresponds to a more predictive variable. 

\begin{figure}[!ht]
    \centering
    \includegraphics[width=\linewidth]{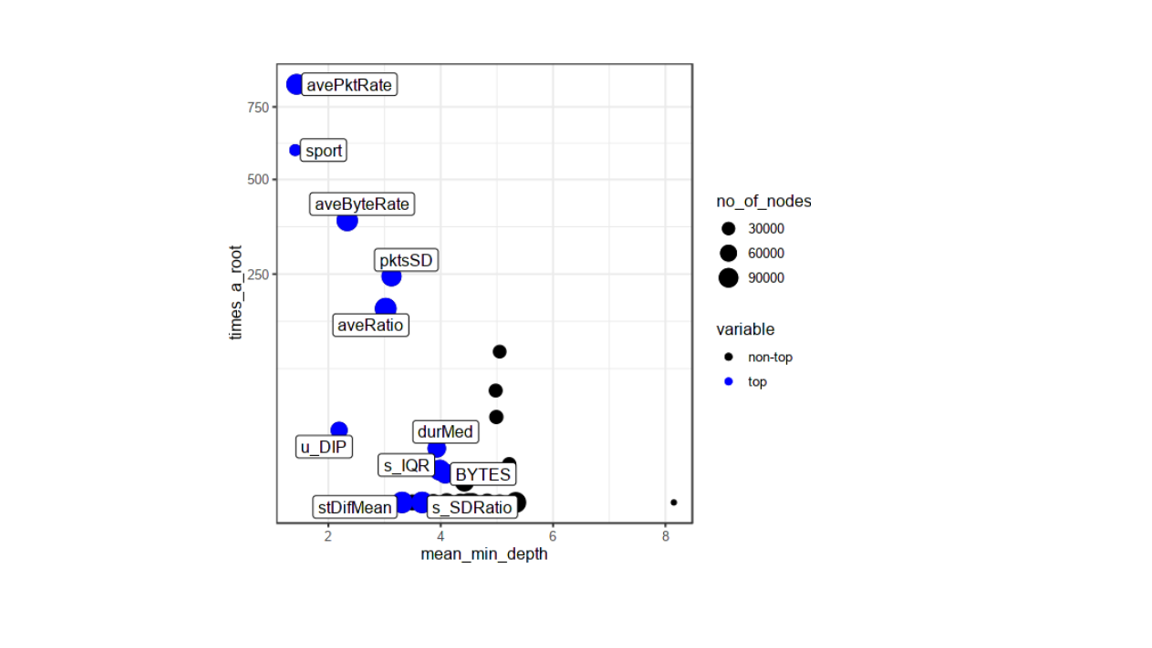}
    \caption{Multi-way Importance plot}
    \label{fig:mwip}
\end{figure}

 To summarize, Paluszynska (2017) focuses on importance measures derived from the structure of trees in the forest. They are the mean depth of first split on the variable, the number of trees in which the root is split on the variable and the total number of nodes in the forest that split on that variable.  We show 3 feature importance scores based on these three importance measurements. In random forest, at each node, a subset of the full set of predictors is evaluated for their strength of association with the dependent variable. The most strongly associated predictor is then used to split on the data. The multi-way importance plot in Figure \ref{fig:mwip} shows that the variables that occur closer to the root are more important, root node being the topmost node in a decision tree. 
The feature importance score \emph{mean min depth}  measures that. If a variable has a low value for \emph{mean min depth}, it implies that on average split occurs closer to the root, therefore is more strongly associated with the dependent variable in each of the bootstrap data subsets. In Figure \ref{fig:mwip}, the mean depth for "avePktRate" is very low, indicating that it is most often the root node in the random forest and therefore very important. The feature importance scores \emph{times a root} and \emph{no of nodes} also measure essentially the same thing: it is the number of trees in which the root is split on the variable and the total number of nodes in the forest that split on that variable respectively. The higher these measures are the more important the feature is for prediction. From Figure \ref{fig:mwip}, we observe that more than 750 trees used the variable "avePktRate" as the top split criterion and over 900,000 nodes used it for splitting. The multi-way importance plot shows the top 11 features ( blue bubbles) out of the feature list of 33 variables used in the model.
\begin{itemize}
    \item We observe 5 out of the 11 variables have high values for \emph{times a root} and \emph{no of nodes} (reflected by the size of the bubble) and low value for the measure \emph{mean min depth}.
    \item The other 6 features still have high values for \emph{times a root} and \emph{no of nodes} .
    \item Features represented in black dots are not important.
\end{itemize}

This is more insightful than just looking at the importance of variables since we are now able to not only assess the relative importance of the variables, but also gain some insight on how important each feature is in an absolute sense. 

\section{The Deep Learning approach}
Deep learning has become a popular vehicle for modeling very large complex streams of unstructured data. There have been a multitude of successful implementations of real world applications involving image classification, text processing, voice recognition, virtual assistants and self driving cars to name a few. Feature engineering with a good model is very difficult and time-consuming to implement on unstructured data. A deep learning network solves this problem.

In general, the Neural Network model is a network of connected neurons. The neurons cannot operate without other neurons - they are connected. Usually, they are grouped in layers and the processed data in each layer is passed forward to the next layers. The last layer of neurons is making decisions. Deep learning is a deep neural network with many hidden layers and many nodes in every hidden layer. Deep learning develops deep learning algorithms that can be used to train complex data and predict the output. The objective here is to test the performance of these models in the context of network security, specifically, using NetFlow to detect botnet attacks.

In addition to the classical ML approach, like decision trees, Random Forest, gradient boosting, etc., deep learning serves as an alternative, which  produces superior performance in the prediction capability though it lacks the explain-ability, which is a trait Machine Learning models. We define $X$, a multivariate NetFlow timeseries matrix $n x p$, defined as the \# of bytes, \# of packets and \# of IP-flows, aggregated at 5 minute intervals for a period of 24 hours. For the matrix $X$, $n$ is the number of IP address traces and $p=288$ is the number of 5-minute bins in a day being studied, $n > p$. Here, we utilize the 1D temporal convolutional neural network to do the prediction, considering the following characteristics of the task:
\begin{enumerate}
    \item The data for our analysis is fixed in length, which has consecutive 288 5-minute time windows;
    \item Every sample has three features for each time point;
    \item Across time, the time series has translation invariance.
\end{enumerate}

The above characteristics of the time series makes 1D temporal convolutional neural network a very competitive choice for the prediction as it can learn the inner representation of the time series.





\subsection{Model Architecture: 1D CNN}
CNN is a class of neural network commonly applied to analyzing visual imagery\cite{krizhevsky2012imagenet}. It has also been applied to image and video recognition\cite{shi2016real}, recommendation systems\cite{ying2018graph}, image classification, medical image analysis, and natural language processing tasks\cite{hu2014convolutional}. Based on the nature of NetFlow as a time series analysis problem, a 1D CNN model is built in order to capture the high dimensional temporal correlations between time intervals. The highly parallel nature of the CNN model enables a fast learning of the model. For the inner structure of the CNN model, dropout \cite{srivastava2014dropout} is used as regularization and ReLU\cite{nair2010rectified}, $Relu(x)  = max(0,x)$, where $x$ is the input value, are used as the activation functions. The cross entropy is treated as the loss function and ADAM is used as an optimizer to train the model. 


\begin{figure}
    \centering
    \includegraphics[width=0.6\linewidth]{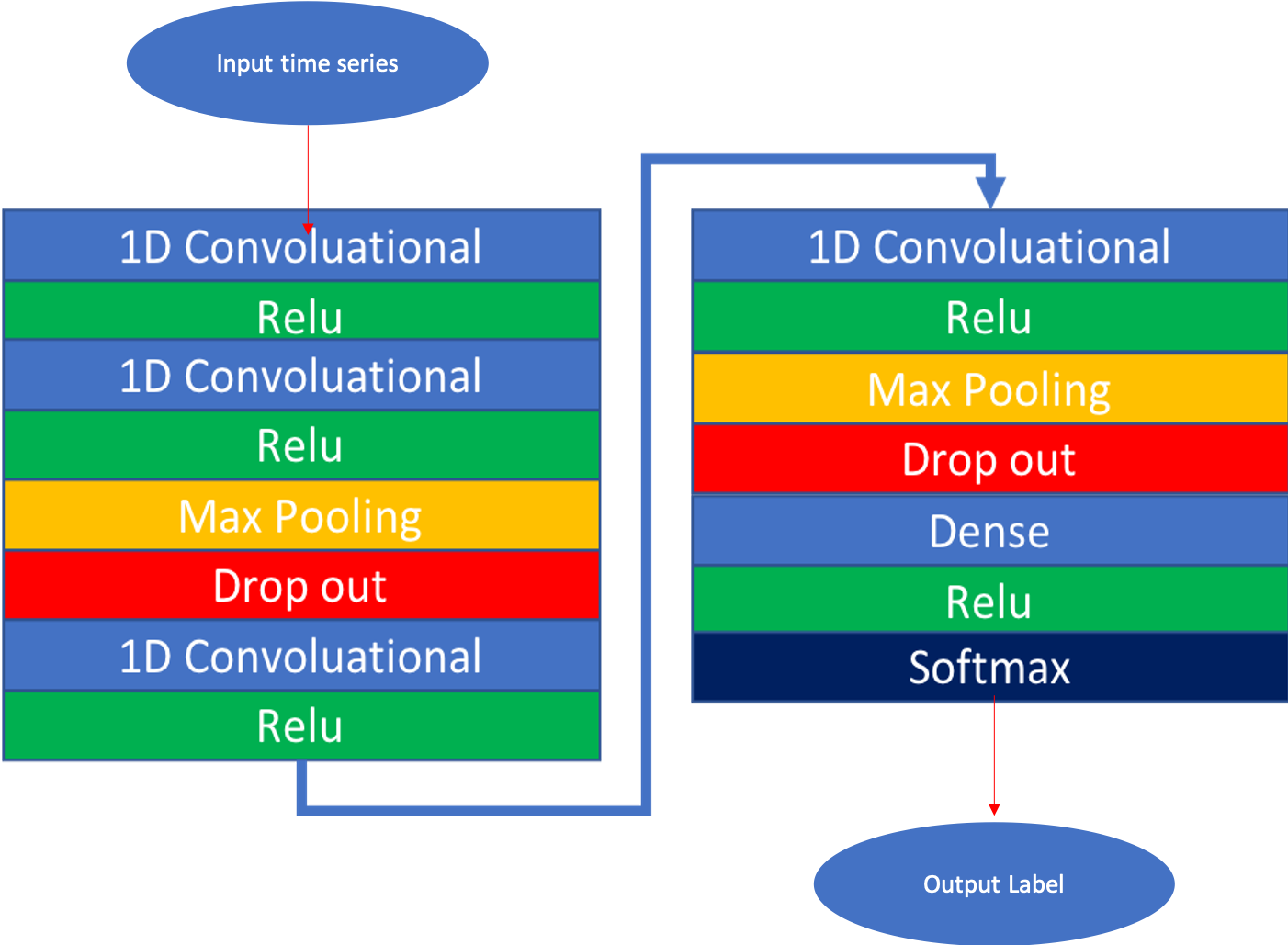}
    \caption{CNN Architectrue.}
    \label{fig:modelArchitecture}
\end{figure}


For our particular application, a total of $4$ convolutional layers are applied, each layer has $32$ filters of size $10$ and stride size $1$. Every filter is of dimension $10\times \textrm{number of previous layers}\times \textrm{number of next layers}$. As shown in Figure \ref{fig:modelArchitecture}, the 1D convolutional kernel takes convolution within each region and stacks the values together. The pooling layer takes the local maximum value for every channel to reduce the amount of parameters and computation in the network, and hence to also control overfitting. The ReLU works as the activation function, which possesses the advantage of sparse activation, better gradient propagation, efficient computation and scale-invariants.


\subsection{Discussion of Results}
\begin{figure}[ht!]
  \centering
  \includegraphics[width=\linewidth]{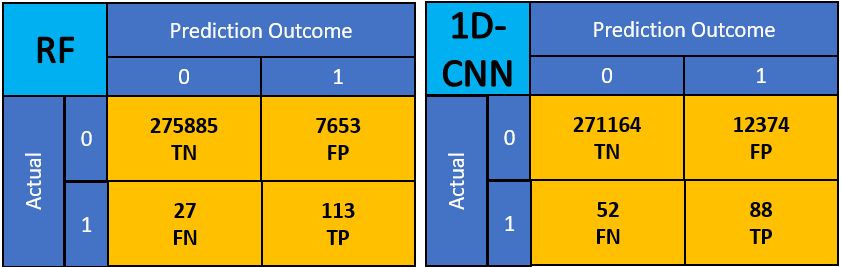}
    \caption{Confusion matrix for Random Forest vs 1D-CNN.}
  \label{fig:confusionMatrix}
\end{figure}

The Random Forest model used hand crafted features developed using the feature exploration step for classification while the convolutional neural network model used the raw multivariate NetFlow timeseries data as input to create its own feature space. One of the advantages to fitting both models is that, the deep learning model with its ability to build its own features can compensate for the features that feature exploration steps, in theory, could have missed. The Random Forest model required a feature extraction step but training was less computationally expensive, while the convolutional neural network model required a GPU server for model training.  Based on the heavily imbalanced nature of our NetFlow applications, in addition to the CNN models, a weighted sampling procedure is applied to the models where the samples from the smaller class are over sampled compared to the samples from the larger class. In Figure \ref{fig:confusionMatrix}, the performance of the Random Forest and 1D temporal Convolutional Neural Network are shown. ). 





\begin{figure}[!ht]
     \centering
     \includegraphics[width=\linewidth]{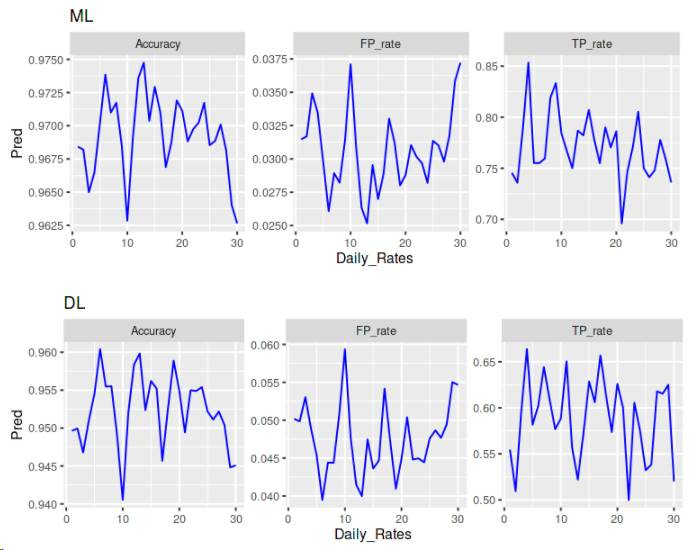}
     \caption{Random Forest Vs CNN: Accuracy, False positive rates and True positive rates for one Month}
     \label{fig:timeComp}
 \end{figure}

Based on the results in Figure \ref{fig:confusionMatrix}, which is the confusion matrix for a single day, both the Random Forest and the CNN models have similar false positive rates (FP),  $0.027$ vs $0.044$, but the true positive rates (TP) are very different. For the convolutional neural network model, the true positive rate is $0.63$ significantly lower than for Random Forest model which is $0.80$.  Figure \ref{fig:timeComp}, shows the daily true positive rates for an entire month of data. For the CNN model, the TP rate fluctuates between 0.52 to 0.67, whereas for the Random Forest model TP rate is between 0.7 and 0.85. This difference could be attributed to the class imbalance. While, for the Random Forest model, down-sampling was implemented to mitigate class imbalance, no mitigation efforts were taken for the CNN model. Currently, the results from both models are used independently as the predicted IP addresses from the two models have only a small overlap. Also, the predicted IP addresses from both models matched IP addresses from external blacklists. Initial analysis shows the Random Forest model matched a much higher percentage of IPs in the blacklists compared to the convolutional neural network model. This was consistent across several months of traffic. If this is used as a metric for comparison, then the Random Forest model outperformed the deep learning CNN model. To understand this phenomenon and other differences, more in depth exploration is needed in the form of comparison of the various statistical attributes of the NetFlow traffic traces for the IPs from the two models.

\section{Conclusion and future work}
From a modeling perspective, the security domain is a wide area with large number of very challenging problems, especially in the area of identification of various malicious events related to attacks like scanning, password guessing, DDoS attacks, malwares, different spams. In this research, we successfully created a statistical framework for identifying botnet traffic, an important threat that is growing in prominence. We constructed two predictive models, one based on Random Forest and the other a weighted CNN, based on deep learning models. Both models ingest daily IP traffic and predict a list of IP addresses (external hosts) that are likely to be malicious command \& control servers. We can identify devices (internal to the network) that are likely to be infected based on the traffic between the malicious IP addresses and the devices. This was based on aggregated traffic between every external host and their respective device traffic without the need to explicitly model traffic between every pair of IP addresses (external host vs devices). The models collectively tried to extract common behaviors of command \& control servers as several botnet families were combined in the training data. To validate the model, the model predicted IP addresses of external hosts were matched with several well established blacklists. The median model predictions for the IPs that matched external lists were high (0.75). In future studies, we will focus on modeling the URLs associated with the traffic between command and control servers and devices. With more data, we can also model individual botnet families. The findings of this research provide an encouraging foundation to further the analysis and prediction of broader malicious events in this challenging field.

\section{Acknowledgement}
We thank Richard Hellstern, and Craig Nohl (AT\&T, CSO) for providing advice, helpful comments and technical expertise related to Network Security.

\bibliographystyle{unsrt}
\bibliography{ref}

\end{document}